\documentclass[10pt]{emulateapj}
\usepackage{apjfonts}

\def\eps{\epsilon}
\def\vareps{\varepsilon}

\shorttitle{Neutrino flux from GRBs} \shortauthors{Li}

\begin{document}

\title{Fermi Limit on the Neutrino Flux from Gamma-ray Bursts }

\author{Zhuo Li\altaffilmark{1,2}}

\altaffiltext{1}{Department of Astronomy and Kavli Institute for
Astronomy and Astrophysics, Peking University, Beijing, China}

\altaffiltext{2}{Key Laboratory for the Structure and Evolution of
Celestial Objects, Chinese Academy of Sciences, Kunming, China}

\begin{abstract}
If gamma-ray bursts (GRBs) produce high energy cosmic rays,
neutrinos are expected to be generated in GRBs via photo-pion
productions. However we stress that the same process also generates
electromagnetic (EM) emission  induced by the secondary electrons
and photons, and that the EM emission is expected to be correlated
to the neutrino flux. Using the Fermi/LAT results on gamma-ray flux
from GRBs, the GRB neutrino emission is limited to be $<20\rm
GeV\,m^{-2}$ per GRB event on average, which is independent of the
unknown GRB proton luminosity. This neutrino limit suggests that the
full IceCube needs stacking more than $130$ GRBs in order to detect
one GRB muon neutrino.
\end{abstract}



\section{Introduction}

The sources of high energy, $>1$PeV, cosmic rays (HECRs) are
expected to produce high energy neutrinos via photo-pion production.
The detection of high energy neutrinos will help to identify the
origin of HECRs. Gamma-ray bursts (GRBs) have long been proposed to
be one of the strong candidates of extragalactic HECR sources
\citep{waxman95}, and was expected to produce high energy neutrinos
\citep{wb97,vietri98,dermer03}. Currently the IceCube, operating in
the full scale, is the most sensitive TeV-scale neutrino telescope,
and is believed to reach the level of GRB neutrino flux. The recent
reported flux limits from IceCube in its uncompleted configuration
have put stringent constraints on GRB neutrinos
\citep{ic22,ic40,ic59}. The observations by IceCube in the full
scale will soon give even more stringent results.

The comparison between the latest non-detection of GRB neutrinos by
IceCube and the positive GRB neutrino prediction challenges GRBs
being the source of HECRs\citep{ic59}\footnote{See, however,
different calculation results in \cite{Li12prd,hummer12,he12}.}.
However, the predicted flux depends on some uncertainties in the GRB
model. First, the neutrino flux is proportional to the proton
luminosity from the GRB jet $L_p$, which is an unknown parameter.
Second, the neutrino flux is proportional to the fraction of proton
energy that is converted into pions $f_\pi$, which further depends
on other uncertain parameters, e.g., the bulk Lorentz factor of the
jet, $\Gamma$, and the radius of the GRB emission regions, $R_{\rm
em}$\citep{wb97,khota06}. Given the fiducial values of these
parameters one can calculate the GRB neutrino flux, as done by
\cite{Guetta04} and IceCube\cite{ic22,ic40,ic59}, but the flux is
subjected to the uncertainties. For example, if the ratio between
energy in accelerated protons and electrons $f_p\equiv L_p/L_e$ is
taken to be $f_p=10$, the predicted neutrino flux by
\cite{ic59,ZhangKumar12} challenge the current non-detection of
neutrinos by the IceCube, but using $f_p=1$ even the photosphere
model of GRBs, which has small $R_{\rm em}$ (hence large $f_\pi$),
survives the current detection limit \citep{Gao12} \citep[see
also][]{murase08,wang09}. A systematic consideration of the
parameter values makes relatively more reasonable
prediction\citep{winter12}, but the result still suffers the
assumptions of parameters.

Here we investigate the neutrino production in GRBs, and emphasize
that the neutrino flux could be related to the electromagnetic (EM)
one in the photo-pion processes \citep{becker10}. Then we derive a
constraint on GRB neutrino flux based on the Fermi observations of
GRB EM emission. As shown below, this normalization of GRB neutrino
flux to gamma-ray emission does not suffers, in particular,  the
uncertainty of $L_p$. Without special mention, we will use in the
following the convention $Q_x=Q/10^x$ and cgs units.

\section{EM and neutrino correlation}

If HECRs are generated in GRB outflows, they interact with the
intense MeV photon field and produce charged pions. The charged
pions decay via the primary mode, $\pi^\pm\rightarrow
\mu+\nu_{\mu}$, and muons further decay via $\mu\rightarrow
e+\nu_e+\nu_\mu$. The initial neutrino flavor ratio in the source is
$\Phi_{\nu_e}^0:\Phi_{\nu_\mu}^0:\Phi_{\nu_\tau}^0=1:2:0$, but due
to neutrino oscillation the flavor ratio observed in distance is
$\Phi_{\nu_e}:\Phi_{\nu_\mu}:\Phi_{\nu_\tau}\approx1:1:1$\citep{particledata}.
The final products induced by one charged pion decay are therefore
four leptons, each equally sharing the original pion energy,
$E_e\approx \vareps_{\nu_e}\approx \vareps_{\nu_\mu}\approx
\vareps_{\nu_\tau}\approx E_\pi/4$.  The generated secondary
electrons and positrons (electrons hereafter for simplicity) may
easily convert their energy into gamma-rays by radiation processes,
e.g., synchrotron and inverse Compton (IC) radiations. Thus there
should be a straightforward relation between the muon neutrino flux
and the gamma-ray flux induced by the secondary electrons,
\begin{equation}\label{eq:1by1}
  F_{\nu_\mu}\approx F_\gamma^{\pi^\pm},
\end{equation}
implying that the neutrino flux may be normalized to and constrained
by the observed gamma-ray flux.

The secondary electrons are very energetic and cool rapidly by
synchrotron radiation since the IC cooling suffers strong
Klein-Nishina effect. Let us calculate the synchrotron photon
energy. For a flat proton distribution with index $p\approx2$
($E_p^2dn_p/dE_p\propto E_p^{2-p}$), and a typical GRB spectrum of a
broken power law with low and high energy photon indices,
$\alpha_\gamma=1$ and $\beta_\gamma=2$, respectively, the generated
pions also show a flat spectrum at $E_\pi\gtrsim
E_{\pi,b}=0.2E_{p,b}$. Here $E_{p,b}$ is the energy of protons that
mainly interact with photons at the spectral-break energy $\eps_b$
\citep{wb97},
$E_{p,b}=1.3\times10^{16}\Gamma^2_{2.5}\eps^{-1}_{b,\rm MeV}\rm eV$
($\eps_{b,\rm MeV}=\eps_b/1\rm MeV$). Moreover at high energies, the
generation of neutrinos and secondary electrons suffers from the
fast cooling of charged pions and muons. Electrons are generated by
muon decays, and the decay time is longer than the muon synchrotron
cooling time at muon energy higher than
$E_{\mu,c}=3\times10^{16}(\eps_e/\eps_B)^{1/2}\Gamma^4_{2.5}\Delta
t_{-2}L_{52}^{-1/2}\rm eV$\citep{wb99,Guetta04,ic22}, where $L$ is
the GRB gamma-ray luminosity, $\Delta t$ is the variability
timescale in GRB light curve, and $\eps_e$ and $\eps_B$ are the
fractions of jet kinetic energy that are carried by accelerated
electrons and magnetic field, respectively. The secondary electrons
are mainly produced by charged pions at
$E_{\pi,b}<E_\pi<(4/3)E_{\mu,c}$. Given $E_e\approx E_\pi/4$, the
synchrotron photon spectrum emitted by the secondary electrons is
mainly dominated by photons between $\eps_b^{(1)}=
6\times10^{12}(\eps_e/\eps_B)^{-1/2}L_{52}^{1/2}\Delta
t_{-2}^{-1}\eps_{b,\rm MeV}^{-2}$eV and $\eps_c^{(1)}=1.5\times
10^{15}(\eps_e/\eps_B)^{1/2}\Gamma_{2.5}^4L_{52}^{-1/2}\Delta
t_{-2}$eV.

High energy photons may interact with the GRB MeV photons and
produce electron/positron pairs, $\gamma\gamma\rightarrow e^{\pm}$.
The $\gamma\gamma$ optical depth exceeds unity for photon energy
above\citep{li10,lw07}
\begin{equation}
  \eps_{\gamma\gamma}\simeq\max(1\Delta
  t_{-2}\Gamma_{2.5}^6L_{52}^{-1},\, 0.3\Gamma_{2.5})\rm GeV.
\end{equation}
Thus $\eps_b^{(1)},\eps_c^{(1)}\gg\eps_{\gamma\gamma}$, i.e., the
synchrotron photons emitted by secondary electrons may be easily
trapped and converted into electron/positron pairs. This initiates
an EM cascade, and the photons only escape once their energy decays
into below $\eps_{\gamma\gamma}$ during the EM cascade. Therefore, a
spectral bump may appear around $\eps_{\gamma\gamma}$, typically
$\eps_{\gamma\gamma}\sim0.1$GeV$-$a few 100's GeV for
$\Gamma\sim100-1000$. One may expect to constrain the GRB neutrino
flux by using the gamma-ray flux of the bump.

\subsection{Additional gamma-ray components}

The observed EM flux, $F_\gamma$, may be contributed not only by the
secondary electrons induced EM cascade radiation
($F_{\gamma}^{\pi^\pm}$), but also by e.g., the leptonic component
from GRB accelerated electrons ($F_\gamma^e$) and neutral pions from
photopion productions ($F_\gamma^{\pi^0}$). Thus the ratio $f\equiv
F_{\nu_\mu}/F_\gamma$, where
$F_\gamma=F_\gamma^{\pi^\pm}+F_\gamma^e+F_\gamma^{\pi^0}$, is
smaller than unity. We consider the $\pi^0$ induced component
$F_\gamma^{\pi^0}$ in details in the following.

There is comparable possibility that the $p\gamma$ interactions
produce neutral pions, which decay fast via
$\pi^0\rightarrow\gamma\gamma$, each photon carrying one half of the
pion energy. These secondary photons consist of a flat spectrum
above
$\eps_{\pi,b}\approx0.5E_{\pi,b}=1.3\times10^{15}\Gamma^2_{2.5}\eps^{-1}_{b,\rm
MeV}$eV. This is well above $\eps_{\gamma\gamma}$, so these photons
also lead to an EM cascade radiation. For very high energy photons
the Klein-Nishina effect will reduce the $\gamma\gamma$ interaction
rates. Typically at photon energy $\eps\ga10$~PeV the $\gamma\gamma$
pair production optical depth becomes smaller than unity, thus
photons can escape if without additional absorption processes.
However, at such high energy the magnetic pair production,
$\gamma+B\rightarrow e^\pm$, becomes important. The absorption
coefficient at photon energy $\eps'$ (comoving frame) for this
process is $k\sim0.1B_5\exp[-600/\eps'_{\rm TeV}B_5]$ cm$^{-1}$. The
optical depth $\tau_{\gamma B}=kR/\Gamma$ is usually larger than
unity at $\eps\ga10$~PeV. Thus the very high energy photons are
still trapped, leading to EM cascade in the emission region.
Assuming that GRB can accelerate protons up to $\sim10^{21}$eV, the
maximum energy of $\pi^0$ induced photons is then
$\eps_{\max}\sim10^{20}$eV. The photons at
$\eps_{\pi,b}<\eps<\eps_{\max}$ will be transferred into EM cascade
radiation.

The energy ratio between the $\pi^\pm$ induced electrons and $\pi^0$
induced photons is
$F_\gamma^{\pi^\pm}/F_\gamma^{\pi^0}\approx{1\over4}\log({4\over3}E_{\mu,c}/E_{\pi,b})/{1\over2}\log(\eps_{\max}/\eps_{\pi,b})$.
The factor 1/4 arises because electrons carry one fourth the
$\pi^\pm$ energy, and 1/2 because the energy ratio of $\pi^0$ to
$\pi^{\pm}$ is $\Phi_{\pi^0}/\Phi_{\pi^\pm}\approx1/2$ by numerical
calculation of photo-pion production in GRBs\citep{winter10}. The
$f$ factor is then calculated as
$f<
F_{\nu_\mu}/(F_\gamma^{\pi^\pm}+F_\gamma^{\pi^0})\approx(1+F_\gamma^{\pi^0}/F_\gamma^{\pi^\pm})^{-1}$.
This is upper limit because the leptonic contribution to gamma-ray
emission is neglected. Plugging the typical values of $E_{\mu,c}$,
$E_{\pi,b}$, $\eps_{\max}$ and $\eps_{\pi,b}$, one gets $f<0.1$.
This number is sensitive to $\Gamma$ (but weakly depends on the
other parameters); for $\Gamma$ varying from $10^2-10^3$ it ranges
from $\sim0.02-0.2$. If the leptonic component is dominant, then
typically $f\ll0.1$.

\subsection{The flux of pile-up bump at $\epsilon_{\gamma\gamma}$}

It should be noted that a significant fraction of the secondary
electrons' energy may be lost due to synchrotron radiation far below
the bump, $\eps\ll\eps_{\gamma\gamma}$. This is the case of strong
magnetic field, $\eps_B\alt\eps_e$. Let us consider the fraction of
energy, $g\equiv F_{\rm bump}/F_\gamma^{\pi^\pm}$, that remains in
the spectral bump at $\eps_{\gamma\gamma}$. The electron energy loss
is dominated by synchrotron radiation for electrons with energy
exceeding $E_{e,\rm KN}=2\times10^{10}\Gamma_{2.5}^2\eps_{b,\rm
MeV}^{-1}$eV, because the IC scattering off the GRB MeV photons
suffers strong Klein-Nishina suppression. So typically $E_{e,\rm
KN}>\eps_{\gamma\gamma}$. Meanwhile, the emitted synchrotron photons
are far below $\eps_{\gamma\gamma}$ for electron energy below
$E_{e,\rm
cut}=8\times10^{12}(\eps_e/\eps_B)^{1/4}\Gamma_{2.5}^2L_{52}^{-1/4}\Delta
t_{-2}^{1/2}(\eps_{\gamma\gamma}/1\,\rm GeV)^{1/2}$eV. Therefore
once the electrons generated during the EM cascade are located at
$E_{e,\rm KN}\lesssim E_e\lesssim E_{e,\rm cut}$, their energy is
leaked out and does not contribute to the spectral bump. For the
first generation electrons, distributed at
$\eps_b^{(1)}<\eps<\eps_c^{(1)}$, this leaking fraction is
$\ln(E_{e,\rm
cut}/\eps_b^{(1)})/\ln(\eps_c^{(1)}/\eps_b^{(1)})\approx0.05$. For
the second generation, $\eps_b^{(2)}<\eps<\eps_c^{(2)}$ with
$\eps_b^{(2)}=
6\times10^{8}(\eps_e/\eps_B)^{-3/2}\Gamma_{2.5}^{-4}L_{52}^{3/2}\Delta
t_{-2}^{-1}\eps_{b,\rm MeV}^{-4}$eV and $\eps_c^{(2)}=3\times
10^{13}(\eps_e/\eps_B)^{1/2}\Gamma_{2.5}^4L_{52}^{-1/2}\Delta
t_{-2}$eV, the leaking fraction is $\ln(E_{e,\rm cut}/E_{e,\rm
KN})/\ln(\eps_c^{(2)}/\eps_b^{(2)})\approx0.55$. The subsequent
generated electrons are not located in the leaking range $E_{e,\rm
KN}\lesssim E_e\lesssim E_{e,\rm cut}$ any more, so that the energy
fraction remained in the spectral bump is typically
$g\approx1-(0.05+0.55)=0.4$. We have assumed fast cooling for
secondary electrons during the cascade, which is valid in a wide
range of parameter values.

Following the same reasoning we also calculate the $g$ values for a
wide range of parameter values. Note, in some cases $E_{e,\rm
KN}<\eps_{\gamma\gamma}$ happens, thus synchrotron dominates the
energy loss in the EM cascade process. In these cases, we simply sum
up for all generations the fraction of synchrotron radiation in the
0.1-300GeV LAT window. As shown in Table \ref{tab}, $g$ is usually
of order unity in the parameter space we concern. The $g$ value
averaged over the parameter space of interests can be conservatively
taken to be $\langle g\rangle>0.1$. As for EM cascade induced by
neutral pion decayed photons, the same reasoning will lead to
similar $g$ value, i.e., on average, $\langle g\rangle>0.1$ also
conservatively holds for neutral pion induced component.

\begin{table}
\caption{The values of factor $g$.}
\begin{center}
\begin{tabular}{cc}
\hline
Parameter values &  $g$ \\
\hline
$\Gamma=100,300,600,1000$ & $0.15,0.42,0.04,0.58$\\
$L/{\rm erg\,s^{-1}}=10^{51},10^{51.5},10^{52.5},10^{53}$ & $0.08,0.23,0.42,0.26$\\
$\Delta t/{\rm s}=10^{-3},10^{-2.5},10^{-1.5},10^{-1}$ & 0.48,0.26,0.19,0.06\\
$\eps_b/{\rm MeV}=0.1,0.3,3$& 0.80,0.24,0.20\\
$\eps_B/\eps_e=0.1,0.3,3$ & 0.31,0.36,0.36\\
\hline
\end{tabular}
\end{center}
\par
\tablecomments{The other parameter without mentioned are taken to be
the typical values: $\Gamma=300$, $L=10^{52}\rm erg\,s^{-1}$,
$\Delta t=10^{-2}$s, $\eps_b=$1 MeV, and $\eps_B/\eps_e=1$, for
which $g=0.42$.} \label{tab}
\end{table}

The detail calculation of the EM cascade emission induced by the
photo-pion production in GRBs has been carried out numerically
\citep{dermer06,asano09,asano10,murase12} to explain the extra
spectral components observed in certain GRBs. The main difference
here is that we estimate in general the contribution of hadronic
components in GRBs.

\section{Fermi limit on neutrino flux}

The Fermi Gamma-ray Space Telescope provided a brand new window to
GRB observations, especially the GeV scale. We constrain the GRB
neutrino flux by Fermi observations.

Fermi satellite has two instruments, Gamma-ray Burst Monitor (GBM),
with wide field of view (FOV) and being sensitive to MeV scale
emission, and Large Area Telescope (LAT), narrower FOV and sensitive
to GeV emission, i.e., from 0.1 to 300 GeV. The LAT has detected
roughly 8\% of the GBM-triggered GRBs that occurred within the LAT
FOV. For these LAT-bright GRBs, the analysis by \cite{Zheng12} has
shown (their Table 2) the photon fluence (i.e., time integrated
photon flux) of each LAT detected GRB in the 0.1-300GeV range. We
then get the average photon flucence for one GRB, $\phi_{\rm
bright}=108\rm ph\,m^{-2}$\footnote{There are totally 22 LAT
detected GRBs listed in \cite{Zheng12}, but we only take the 15
reported by Fermi collaboration and neglect the other 7 discovered
by using the special method}. Since there is quite few photons
detected above few 10's GeV, the photon number in 0.1-10GeV range is
practically equal to that in the 0.1-300GeV range. For the LAT-dark
GRBs, the upper limits of LAT detection are given first by \cite{lat
limit1,lat limit2}, but we here use the results from Fermi team. In
their recent paper \citep{lat-dark} the Fermi team has analyzed the
LAT-dark GRBs in roughly 3 years operation, which are the other 92\%
GBM-triggered GRBs that are in the LAT FOV. The upper limits to the
gamma-ray flux in 0.1-10GeV range for each GRB has been listed in
their Table 1. We use the upper limits in the last column to
calculate the average upper limit of all these LAT-dark events,
$\phi_{\rm dark}=34.6\rm ph\,m^{-2}$ (0.1-10GeV range). Let us
regard $\phi_{\rm bright}$ as the upper limits to the bright GRBs,
then we can calculate the average upper limit to all GRBs, including
both bright and dark GRBs, as $\phi_{\rm limit}=8\%\times\phi_{\rm
bright}+92\%\times\phi_{\rm dark}=40\rm ph\,m^{-2}$ in 0.1-10GeV
range. Assuming a flat photon spectrum,
$\eps^2dn_\gamma/d\eps\propto\eps^{2-\gamma}$ with $\gamma=2$, the
average limit to the energy fluence per GRB is $
  F_{\gamma,\rm limit}=20\rm GeV\,m^{-2}
$ in 0.1-10GeV range. For softer photon spectrum with $\gamma=3$,
the limit becomes $F_{\gamma,\rm limit}=8.5\rm GeV\,m^{-2}$, smaller
by a factor of 2.3, whereas for harder spectrum with $\gamma=1.5$,
$F_{\gamma,\rm limit}=75\rm GeV\,m^{-2}$, larger by 3.7. Given the
correlation between neutrino flux and gamma-ray flux, the average
neutrino fluence per GRB is
\begin{equation}\label{eq:limit}
  F_{\nu_\mu}<\langle f/g\rangle F_{\gamma,\rm limit}=20\langle f/g\rangle\rm GeV\,m^{-2}.
\end{equation}
Note, $\langle f/g\rangle \la1$ is estimated in the previous
section.

\section{Neutrino detection rate}

Besides the normalization of neutrino flux, one still needs the
neutrino spectral form in order to calculate the neutrino detection
rate by experiments, like the IceCube. For a GRB spectrum with
photon indices $\alpha_\gamma$ and $\beta_\gamma$, and for a flat
proton distribution with $p=2$, the GRB neutrino spectrum generated
by the $p\gamma$ interactions can be approximated as
$dn_\nu/d\vareps\propto \vareps^{-\alpha_\nu}$ at
$\vareps<\vareps_1$, $dn_\nu/d\vareps\propto\vareps^{-\beta_\nu}$ at
$\vareps_1<\vareps<\vareps_2$, and
$dn_\nu/d\vareps\propto\vareps^{-\gamma_\nu}$ at $\vareps_2<\vareps$
(For simplicity hereafter we take $\vareps\equiv\vareps_{\nu_\mu}$)
where the spectral indices are $\alpha_\nu=3-\beta_\gamma$,
$\beta_\nu=3-\alpha_\gamma$, and $\gamma_\nu=\beta_\nu+2$, and the
break energies are $\vareps_1=E_{\pi,b}/4$ and
$\vareps_2=E_{\mu,c}/3$. The normalization of the neutrino flux is
obtained by the requirement
$\int_0^\infty\vareps(dn_\nu/d\vareps)d\vareps=\langle f/g\rangle
F_{\gamma, \rm limit}$. For typical GRB spectrum with
$\alpha_\gamma=1$ and $\beta_\gamma=2$, the specific neutrino flux
at $\vareps=\vareps_1$ is
$dn_\nu/d\vareps(\vareps=\vareps_1)=\langle f/g\rangle F_{\gamma,\rm
limit}/\vareps_1^2[(3/2)+\log(\vareps_2/\vareps_1)]$. Note, the
broken power law form is a good approximation to the spectral
profile of GRB muon neutrinos from a full numerical calculation with
the effect of neutrino mixing \citep[see, e.g.,][]{winter11}.

Given the effective area of the neutrino experiment, the average
neutrino number that can be detected in one GRB can be calculated as
$N_{\rm det}=\int_0^\infty A_{\rm eff}(dn_\nu/d\vareps)d\vareps$.
For the neutrino spectrum with $F_{\nu_\mu}<20\langle f/g\rangle\rm
GeV\,m^{-2}$, $\varepsilon_1=7\times10^{14}$eV and
$\varepsilon_2=10^{16}$eV, the detection rate by the IceCube
40-string configuration (using the effective area averaged over
neutrino incident angles) is $N^{\rm IC40}_{\rm
det}=2.5\times10^{-3}\langle f/g\rangle$ (for $\gamma=2$) per GRB,
so more than $1/N^{\rm IC40}_{\rm det}=400\langle f/g\rangle^{-1}$
GRBs are needed in the stacking analysis in order to detect one GRB
muon neutrino. For the full scale IceCube with larger effective area
$A^{\rm IC86}_{\rm eff}\approx3A^{\rm IC40}_{\rm
eff}$\citep{Karle10}, and for $\langle f\rangle<0.1$ and $\langle
g\rangle>0.1$, one needs to stack $>130$ GRBs to detect one GRB muon
neutrino (or $>40$ and $>310$ for $\gamma=1.5$ and 3, respectively).

\section{Conclusion and discussion}

We show that in the $p\gamma$ processes there is a correlation
between the generated EM (electrons and photons) and neutrino fluxes
\citep[see, also,][]{bg12}. The EM radiation cascades down to
typically GeV scale before escaping from the GRB emission region.
Using the Fermi/LAT observations the average neutrino flux per GRB
is constrained to be below $20\rm GeV\,m^{-2}$, implying that the
IceCube in its 40 string configuration needs $>400$ GRBs to detect
one muon neutrino. This suggests that the GRB samples analyzed by
the IceCube collaboration in \cite{ic59} may be not large enough. As
for the full scale IceCube, the stacking of $>130$ GRBs is required.

The derivation of neutrino flux using the EM-neutrino correlation is
less dependent on the uncertainties in GRB models. Especially,
unlike the previous calculations, the approach here is completely
independent of the poorly known $L_p$ (or $f_p$). Yet the derivation
here is not parameter free. The location $\eps_{\gamma\gamma}$ where
EM cascade radiation significantly piles up depends strongly on
$\Gamma$. However, based on the theory of internal shock model for
GRBs a robust bound $10^2<\Gamma<10^3$ is obtained\footnote{The fact
that the internal shock radius $R_{\rm coll}\approx\Gamma^2c\Delta
t$ is not within the photospheric radius $R_{\rm
ph}=\sqrt{\sigma_T\Delta N_e/4\pi}$, with $\Delta N_e=L_k\Delta
t/\Gamma m_pc^2$, requires $\Gamma>10^2(L_{52}/\Delta
t_{-2})^{1/5}$. On the other hand, the initial fireball of GRBs can
only be accelerated by the thermal pressure up to
$\Gamma<10^3(L_{52}/r_{0,7})^{1/4}$, where $r_0$ is the source size
\citep[see, e.g][]{Waxman rev}.}. Moreover, observations also
constrain $\Gamma$ in a similar narrow range
\citep{afterglowLF,afterglowLF2,afterglowLF3,afterglowLF4,peer07,li10}.
Therefore for $\Gamma\approx10^2-10^3$ the spectral bump lies right
in the energy window of the LAT. Furthermore, the factor $f/g$ in eq
(\ref{eq:limit}) depends on parameters. Thus we scan a wide range of
parameter space of interests (e.g., Table \ref{tab}), and find out
$\langle f/g\rangle<1$ is a conservative limit in a statistic sense.

Some comments on the derivation are made here. (1) It is assumed
that the $p\gamma$ interactions happen at radius
$R\approx\Gamma^2c\Delta t$, which may not hold in the other GRB
models. Thus one should remind that the neutrino flux is model
dependent \citep{ZhangKumar12}. (2) We neglect the jet expansion
effect, since $\eps_{\gamma\gamma}$ increases with $R$ \citep{li10},
i.e. the expansion does not help in delaying the escape of photons.

By normalizing the neutrino flux to the observed flux of ultrahigh
energy, $>10^{19}$eV, cosmic rays (UHECRs), \cite{wb97,wb99} derived
the diffuse GRB neutrino flux,
$\vareps^2\Phi_\nu=1.5\times10^{-9}(f_\pi/0.2)\rm
GeV\,cm^{-2}s^{-1}sr^{-1}$ ($\vareps_1<\vareps<\vareps_2$). Since
GBM observations show roughly $R_{\rm GRB}\sim10^3$ GRBs per year
for full sky, the average neutrino fluence per GRB is
$F_\nu=7.5(f_\pi/0.2)(R_{\rm GRB}/10^3{\rm yr^{-1}})^{-1}\rm
GeV\,m^{-2}$, smaller by about a factor of 3 than the limit of
eq.(\ref{eq:limit}). If those GRBs that do not trigger GBM
contribute significant fraction of the diffuse neutrino flux, the
average neutrino flux per triggered GRB is even lower. Using the
average Fermi limit and assuming that the GRB detection rate is
$R_{\rm GRB}=10^3\rm yr^{-1}$ and that the untrigger-GRB
contribution is not important, the full-scale IceCube, covering half
sky, detects $\alt4$ muon neutrinos associated with GRBs each year.

Given the neutrino limit, one can constrain the HECR production in
GRBs, i.e., $f_p$. It is derived that the protons with energy
$E_{p,b}$ typically loses a fraction $f_{\pi,b}\sim0.2$ of its
energy by $p\gamma$ interactions\citep{wb97}, and that the
luminosity ratio of neutrinos to protons is\citep{Li12prd}
$$
  \frac{L_{\nu_\mu}}{L_p}\approx\frac18f_{\pi,b}\frac{\log({4\over3}E_{\mu,c}/E_{\pi,b})}{\log(E_{p,\max}/E_{p,\min})}
  \sim\frac{f_{\pi,b}}{40},
$$
with $E_{p,\max}$ ($E_{p,\min}$) the maximum (minimum) energy of
accelerated protons. The proton to electron ratio is given by
$f_p=L_p/L_{\gamma,\rm
MeV}=(L_p/L_{\nu_\mu})(F_{\nu_\mu}/F_{\gamma,\rm
MeV})<13(F_{\gamma,\rm MeV}/3\times10^{-5}{\rm
erg\,cm^{-2}})^{-1}(f_{\pi,b}/0.2)^{-1}$, where $F_{\gamma,\rm MeV}$
is the average fluence of the GBM detected GRBs in the MeV domain.
On the other hand, the wide energy range observations by Fermi have
made more straightforward constraint on $f_p$. Fermi show that the
flux ratio of LAT to GBM is typically $L_{\gamma, \rm
GeV}/L_{\gamma, \rm MeV}\lesssim0.1$\citep{latratio}. Given
$L_{\nu_\mu}/L_{\gamma,\rm GeV}=f/g <1$, we have
$f_p=L_p/L_{\gamma,\rm
MeV}=(L_p/L_{\nu_\mu})(L_{\nu_\mu}/L_{\gamma,\rm GeV})(L_{\gamma,\rm
GeV}/L_{\gamma,\rm MeV})\lesssim20(f_{\pi,b}/0.2)^{-1}$, similar to
the above constraint derived from neutrino limit. Surely, both
constraints depend on the uncertain $f_{\pi,b}$. It has been noticed
that $f_p\gtrsim10$ is required to explain the observed UHECRs as
GRB origin\citep{CRfp,CRfp2,CRfp3}. Combining the neutrino (or
gamma-ray) and UHECR constraints, the allowed range of $f_p$ is then
quite small.

There may be some caveats that the neutrino limit in this paper can
be avoided in some cases. First, we have used the GeV scale, 0.1-few
100's GeV, flux to constrain the neutrino flux, but it could be that
the cutoff photon energy is much larger, e.g.,
$\eps_{\gamma\gamma}\gg100$ GeV, so the observations in $<100$ GeV
range do not make sense. Second, if for unknown reasons it happened
that the generated secondary electrons, i.e., induced by the charged
pion decay or $\gamma\gamma$ pair production, do not radiate at all,
the EM-neutrino correlation would not exist. Finally, it worths
noting that neutrinos may be delayed or anticipated respect to the
GRB photons \citep[see, e.g.,][]{guetta13}. In this case the IceCube
analysis should be done in a larger window time.

\acknowledgments

The author thanks referees for constructive comments. This work is
partly supported by the NSFC (11273005), the MOE Ph.D. Programs
Foundation, China (20120001110064) and the CAS Open Research Program
of Key Laboratory for the Structure and Evolution of Celestial
Objects.

\emph{Note added in proof.} The IceCube non-detection of neutrinos
from the recent low-redshift LAT-detected GRB 130427a
(\verb"http://gcn.gsfc.nasa.gov/gcn3/14520.gcn3") is consistent with
the EM-neutrino correlation. The LAT fluence of $\sim10^4$erg
cm$^{-2}$ only implies a muonic neutrino event number of about <0.4
in IceCube.

\end{document}